\documentclass[twocolumn,prl,aps,showpacs,preprintnumbers,superscriptaddress,amsmath,amssymb]{revtex4}

\usepackage{graphicx}

\newcommand{\ybgion}{\textsuperscript{172}Yb\textsuperscript{+}}

\newcommand{\imagewidth}{0.85\columnwidth}

\begin{document}


\title{Individual addressing of trapped ions and coupling of motional and spin states using rf
radiation}

\author{M. Johanning}
 \affiliation{Fachbereich Physik, Universit\"at Siegen, 57068 Siegen, Germany}

\author{A. Braun}
 \affiliation{Fachbereich Physik, Universit\"at Siegen, 57068 Siegen, Germany}

\author{N. Timoney}
 \affiliation{Fachbereich Physik, Universit\"at Siegen, 57068 Siegen, Germany}

\author{V. Elman}
 \affiliation{Fachbereich Physik, Universit\"at Siegen, 57068 Siegen, Germany}

\author{W. Neuhauser}
 \affiliation{Institut f\"ur Laser-Physik, Universit\"at Hamburg, Luruper Chaussee 149, 22761 Hamburg, Germany}

\author{Chr. Wunderlich}
 \affiliation{Fachbereich Physik, Universit\"at Siegen, 57068 Siegen, Germany}

\date{\today}

\begin{abstract}
Individual electrodynamically trapped and laser cooled ions are
addressed in frequency space using radio-frequency radiation in the
presence of a static magnetic field gradient. In addition, an
interaction between motional and spin states induced by an rf field
is demonstrated employing rf-optical double resonance spectroscopy.
These are two essential experimental steps towards realizing a novel
concept for implementing quantum simulations and quantum computing
with trapped ions.
\end{abstract}



\maketitle

Quantum simulations addressing a specific scientific problem and
universal quantum computation are expected to yield new insight into
as of yet unsolved physical problems that withstand efficient
treatment on a classical computer (e.g., \cite{Feynman1982}).
Already a small number of qubits (i.e., a few tens) used for quantum
{\it simulations} could solve problems even beyond the realm of
quantum information science. Creating and investigating entanglement
in large physical systems is a related important experimental
challenge with implications for our understanding of the transition
between the elusive quantum regime and the classical world
\cite{Dur2004}.

Laser cooled atomic ions confined in an electrodynamic cage have
successfully been used for quantum information processing (QIP)
\cite{Schmidt-Kaler2003} and advantages and difficulties associated
with this system have been and still are subject to detailed
investigations. The electromagnetic radiation used to coherently
drive ionic resonances that serve as qubits needs to be stable
against variations in frequency, phase, and amplitude over the
course of a quantum computation or simulation. Experimentally this
is particularly challenging when laser light is used for realizing
quantum gates. When employing laser light additional issues need to
be dealt with to allow for accurate qubit manipulation such as the
intensity profile of the laser beam, its pointing stability, and
diffraction effects. Furthermore, the motional state of the ion
chain strongly affects the gate fidelity which requires ground state
cooling and low heating rates during the gate operation
\cite{Wineland1998}. Also, spontaneous scattering of laser light off
excited electronic states may pose a limit for the coherence time of
a quantum many-body state. The probability for scattering can be
reduced by increasing the detuning from excited states (when two
laser light fields are used that drive a Raman transition between
hyperfine or Zeeman states) which, however, leads to an increasing
demand for laser power \cite{Steane2007}.

For generating Raman laser beams with a desired frequency
difference, first a radio-frequency (rf) or microwave signal at this
difference frequency has to be generated that is then "imprinted" on
the laser light and send to the ions. Using rf or microwave
radiation directly for coherent driving of qubit transitions is
impeded in usual ion trap schemes, since, (i) individual addressing
of qubits by focusing radiation on just one ion is difficult due to
the long wavelength of rf radiation, and (ii) the required coupling
between qubit states and motional states (that serve as a quantum
bus \cite{Cirac1995}) as measured by the Lamb-Dicke parameter is
negligibly small. Here, we demonstrate individual addressing of
trapped ions in frequency space instead of position space using rf
radiation, thus avoiding the technical and fundamental difficulties
mentioned above that are associated with the use of laser light for
coherent manipulation of trapped ions.

This is achieved by applying a spatially varying magnetic
field such that the relevant Zeeman states of each ion exhibit a
site-specific resonance frequency. Also, this magnetic gradient
field mediates coupling between Zeeman states and motional states
\cite{Mintert2001}. Such a coupling is required, for instance, for
conditional quantum dynamics (quantum gates) with two or more ions.
It may be also used for any experiment where coupled dynamics of
motional and spin states is desired, for instance, sideband cooling
of trapped atoms or ions \cite{Wunderlich2005}.

At the same time, these are two fundamental experimental steps
undertaken on the way towards realizing a novel physical system for
quantum information science, the ion spin molecule. The term ion spin
"molecule" is used here to describe a pseudomolecule
\cite{Wineland1987} exposed to a magnetic field gradient that
induces pairwise coupling $J_{ij}$ between the internal qubit states
(spins) of ion $i$ and $j$ ($i,j=1,2,3,...,N$ with $N$ the number of
ions) \cite{Wunderlich2002,Wunderlich2003}.
An ion spin molecule is useful for quantum simulations, since spin
chains with globally and locally adjustable coupling parameters can
be realized and are well suited, for instance, for simulating
magnetism of quantum systems \cite{Schatz2008}. Also, ion spin
molecules may be used  to implement a classical neural network or
error-resistant quantum computing based on a quantum neural network
\cite{Pons2007}.

The experiments presented here were performed with laser cooled
\ybgion-ions in a linear Paul trap. A detailed description of the
trap and the experimental set-up is given in \cite{Balzer2006}.
The electric dipole transition between the S$_{1/2}$ ground state and
the P$_{1/2}$ excited state near 369~nm is used for Doppler cooling
and state-selective detection by detecting resonance fluorescence
with a photomultiplier or an intensified CCD camera ("cooling
fluorescence"). Individual addressing of ions and the coupling between motional and
internal ionic states is demonstrated using rf-optical double
resonance spectroscopy on the metastable D$_{3/2}$-state (lifetime
of 52.2~ms \cite{Gerz1988}) populated through spontaneous decay from
the P$_{1/2}$ level. The degeneracy of its Zeeman-manifold
is lifted by a magnetic field $B=B_0 + z \;\partial_z\,B $ composed
of an offset field $B_0$ and an additional field with constant
gradient $\partial_z\,B$ ($\vec{z}$ is the axis of rotational
symmetry of the trapping potential and thus of the linear ion
string). The magnitude of $B$ determines the resonance frequency of
magnetic dipole transitions between Zeeman states. The (linear)
Zeeman shift $\Delta E_J$ by a magnetic field $B$ is given by
$\Delta E_J = g_J \, m_J \, \mu_B \, B$ with Landé g-factor $g_J$,
magnetic quantum number $m_J$, and the Bohr magneton $\mu_B$.
Magnetic dipole transitions between levels with $\Delta\, m_J =
\pm1$ with resonance frequency
$f = g_J \, \mu_B \, B\, /\, h$
are driven using an rf field that is generated by a dipole coil. The
magnitude of $B_0$ was chosen for most experiments to be about
0.67~mT resulting in a resonance frequency $f \approx 7.5$~MHz.

This four level system can be initialized by optical pumping: a
laser light field near 935~nm \cite{{Bell1991}} (labeled "repumper"
in what follows) with its linear polarization aligned parallel to
$\vec{B}$ allows $\pi$-transitions ($\Delta m=0$) only. Since
repumping occurs through the state $[3/2]_{1/2}$ with $m = \pm 1/2$,
population accumulates in the "dark" $m = \pm 3/2$ states, and in
turn the resonance fluorescence near 369 nm stops. Upon coupling the
Zeeman sublevels with a resonant rf field the resonance fluorescence
reappears. Modelling the ionic dynamics in the stationary state
using rate equations shows the intensity of scattered light near 369
nm to be proportional to the population in the D$_{3/2}$ Zeeman
states with $m = \pm 1/2$. The stationary state is obtained when the
cooling laser, the repumper and the rf field are applied
simultaneously and the magnetic field is aligned parallel to the
polarization of the repumper to allow for optical pumping. Then
rf-optical double resonance spectra are observed. When the rf is
scanned across this resonance frequency Lorentzian line shapes are
observed whose linewidth is determined by the intensity of both, the
rf radiation and the laser light near 935 nm (repumper). We observed
linewidths between 15~kHz and 310~kHz for rf powers between 25~mW
and 8~W and repumping powers between .3 and 3.8~$\mu$W focussed to a
spot size of approximately 100~$\mu$m in diameter. For a vanishing
gradient $\partial_z B$, all ions have the same resonance frequency
$f$ determined by the offset field $B_0$.

Addressing of individual ions is achieved without focussing the
radiation down to less than the spatial separation of adjacent ions
by now applying a spatially varying magnetic field $B=B_0 + z
\;\partial_z\,B$ such that the Zeeman shift is different for each
ion. The frequency separation of the resonances of two neighboring
ions is given by
\begin{equation}
\Delta f = g_J \, \mu_B \, \delta z \, \partial_z B\,/\,h
\label{eq:freqsep}
\end{equation}
with ion separation $\delta z$. The inhomogeneous field along the
trap axis is created by cylindrical Nd-permanent-magnets with a
diameter of 30~mm in quadrupole configuration separated by
approximately 12~cm. The magnetic flux density at the surface of the
permanent magnets exceeds 1~T.
\begin{figure}[t]
    \centering
        \includegraphics[width=\imagewidth]{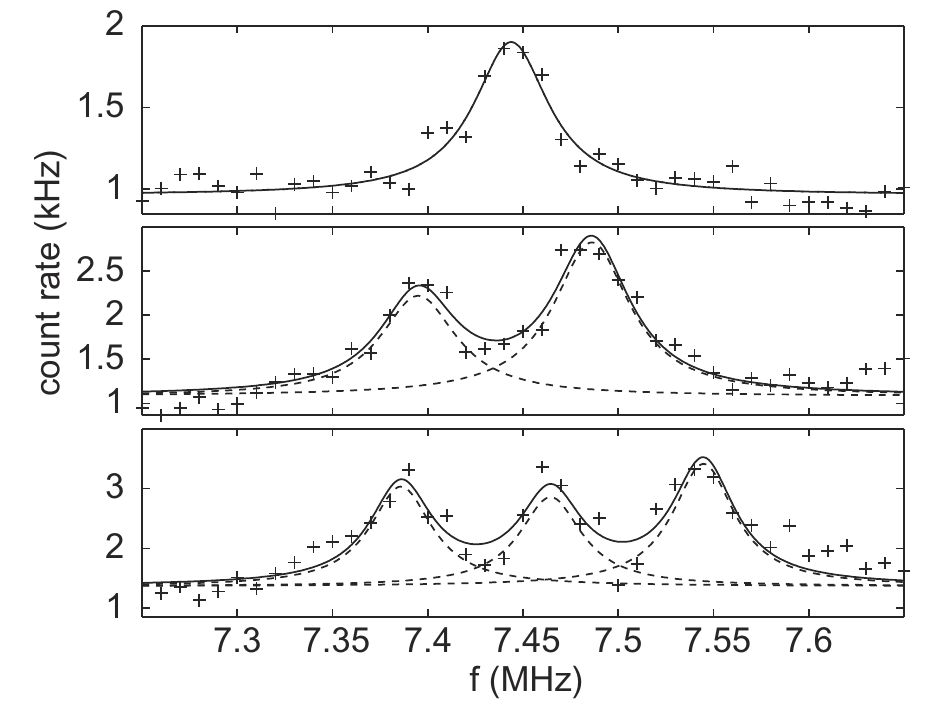}
\caption{Individual addressing in frequency space of one (top), two
(middle) and three (bottom) trapped ions, respectively. The
frequency separation between neighboring ions is extracted by
fitting a sum of Lorentzian line profiles and is $91(3)$~kHz for two
ions and $79 (4)$~kHz (middle-left) and $80 (4)$~kHz (middle-right)
for three ions. The average ratio of these frequency separations is
$1.15 (7)$ which is in good agreement with the analytical prediction
of $\sqrt[3]{8/5} \approx 1.17$ for a constant gradient. Together
with the extracted FWHM of $49 (5)$~kHz the spurious excitation of
adjacent ions is $6.7 (8)$~\% in the case of two ions.}
 \label{fig:addressingpm}
\end{figure}

\begin{figure}[t]
    \centering
        \includegraphics[width=\imagewidth]{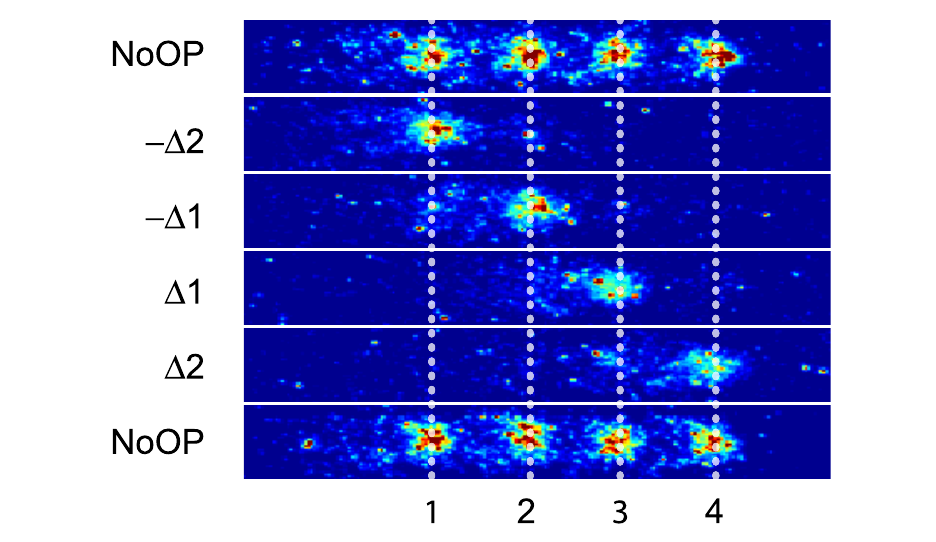}
\caption{Individually addressing a particular $^{172}$Yb ion that is
part of a linear Coulomb crystal composed of four $^{172}$Yb ions.
Six images of the same ion string are shown. These were taken
successively from top to bottom. The uppermost and lowermost image
labeled 'NoOP' are recorded with no optical pumping, that is, all
ions scatter light. For the intermediate pictures, the magnetic
field direction is rotated to coincide with the polarization of the
repumper, allowing optical pumping. In addition, the rf frequency
was set to the expected resonance of one particular ion. The
detuning $\Delta1$ or $\Delta2$ of the rf frequency necessary to
address a desired ion is indicated on the left-hand-side of the
graph and is given relative to the Zeeman resonance of a single ion.
Here, $\Delta1$ = 26.3~kHz and $\Delta2$ = 83.2~kHz. Here the amount
of spurious excitation of adjacent ions is comparable to the crosstalk
visible in figure \ref{fig:addressingpm}.}
\label{fig:addressingccd}
\end{figure}

When scanning the rf frequency, one ion after the other comes into
resonance and consequently scatters light. The resonance fluorescence spectra in figure \ref{fig:addressingpm} show individual addressing of strings of up to three ions with well resolved peaks each corresponding to a resonance of an individual trapped ion.
The axial trap frequency was measured by parametric heating to be
$36.3(5)$~kHz, yielding ion separations of $31.7 (3)$~$\mu$m for two
ions and $25.0(2)$~$\mu$m for three ions, respectively. (The radial
trap frequency was $\approx$ 600 kHz.) The experimentally determined
frequency separation of the ionic resonances (figure
\ref{fig:addressingpm}) together with equation (\ref{eq:freqsep})
yield the magnitude of the field gradient $\partial_z
B=0.27(1)$~T/m. The maximum gradient observed was $0.51 (2)$~T/m.
Thus frequency separations of up to 130~kHz could be achieved, and,
with a line width of 15~kHz, unwanted excitation of vicinal ions is
reduced to below $0.4~\%$.
Figure \ref{fig:addressingccd} is composed of spatially resolved
images showing a string of four ions taken with an intensified CCD
camera while setting the rf frequency to specific values that
correspond to individual ionic resonances. It demonstrates that the
peaks of the resonance fluorescence observed in figure
\ref{fig:addressingpm} indeed originate from different ions.
Moreover, it shows that single ions are selectively repumped by
setting the rf frequency to a particular individual ionic resonance.
Here, the spatial separation of the outer ions is 61.3~$\mu$m (trap
frequency of $2\pi\times 46$~kHz). Using equation (\ref{eq:freqsep})
this yields a gradient of 0.24~T/m for this experiment.

When a trapped ion interacts  with electromagnetic radiation, this
may not only affect the atom's internal state but also its motional
state evidenced by motional sidebands in an ionic excitation
spectrum \cite{Wineland1979}.
The coupling between internal and motional dynamics of the ion is
exploited in ion traps to realize conditional quantum dynamics, for
instance, a CNOT gate \cite{Cirac1995,Monroe1995,Schmidt-Kaler2003}.
The Lamb-Dicke parameter, $\eta$ determines its strength:
$\eta =  \sqrt{\hbar^2 k^2 / 2 m \hbar \nu_1}$.
Here, the wave number $k= 2\pi / \lambda$ where $\lambda$ is the
wavelength of the radiation exciting the ion, $m$ the ion's mass,
and $\nu_1$ the secular vibrational frequency. $\eta$ is negligibly
small when using rf or microwave radiation in contrast to optical
fields. For the interaction in an inhomogeneous magnetic field,
however, the Zeeman energy and thus the ion's equilibrium position
may become state dependent.
The coupling of internal and motional state is now described by an
{\it effective} Lamb-Dicke parameter $\eta_{\text{eff}}$, that, for a
single ion, is given by the ratio between the frequency change of
the resonance over the spatial extension $\Delta z =
\sqrt{\hbar/2m\nu_1}$ of the ion's wavefunction and the trap
frequency \cite{Mintert2001}:
$|\eta_{\textrm{eff}}| = \sqrt{\eta^2 + \kappa^2} \approx \left|
\kappa \right| = \left|\frac{\Delta z \,\partial_z f_{\rm
rf}}{\nu_1} \right|$
Here, the experimentally determined parameters for the magnetic
field gradient and trap frequency yield $\eta_{\text{eff}} =
1.1\cdot10^{-3}$.

Neglecting saturation, the amplitudes of the lower and upper
motional sidebands have a relative height compared to the carrier
given by
\begin{equation}
    \frac{a_l}{a_0} = \left< n\right> \eta_{\text{eff}}^2 \quad \text{and}  \quad \frac{a_u}{a_0} = (\left< n\right>+1)\eta_{\text{eff}}^2
    \label{eq:sidebands}
\end{equation}
with $\left< n\right>$ being the mean phonon number
\cite{Wineland1979}, and for large $\left< n\right>$ we have $a_l
\approx a_u \equiv a_s$.

\begin{figure}[t]
    \centering
        \includegraphics[width=\imagewidth]{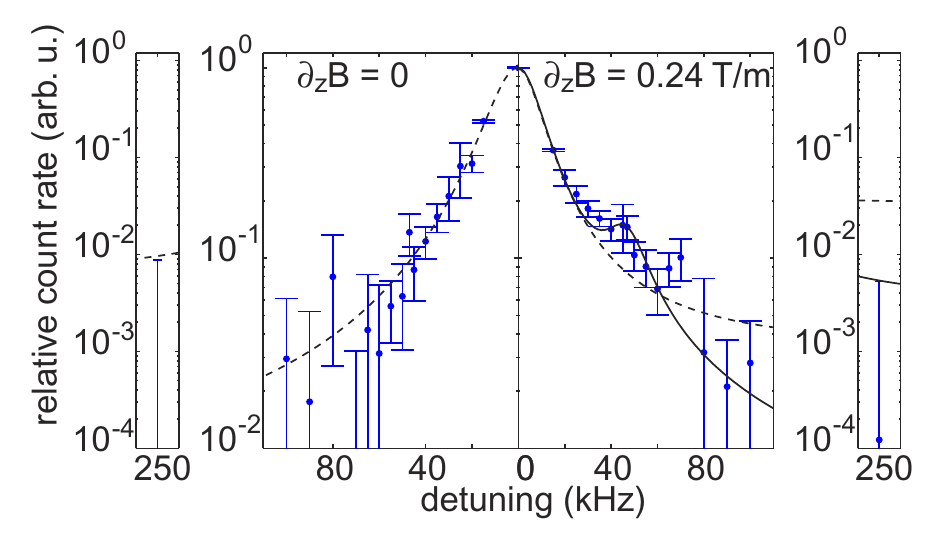}
\caption{Comparison of two spectra
recorded with a single ion exposed to a magnetic field
gradient (right-hand side) and without field gradient (left-hand
side), respectively. The solid line represents a fit using two
Lorentzian lines (rhs), the dashed lines represent fits using a
single Lorentzian (lhs and rhs). The motional sideband accompanying the spin resonance
signifies coupling between spin degrees of freedom and the motion of
the trapped Yb$^+$ ion. }
 \label{fig:comparecoupling}
\end{figure}

The spectra shown in figure \ref{fig:comparecoupling} differ only in
the magnetic field $B$ applied to the ion: $B$ is homogeneous for
the left spectrum (i.e. $\partial_z B_1 =0$ and thus
$\eta_{\textrm{eff}} =0$), while a field gradient is applied for the
right spectrum. The measurement procedure is in principle as
detailed above, with alternating rf-optical double resonance (total
duration of 150 ms) and cooling cycles (duration of 100 ms).
However, in order to resolve the motional sideband with sufficient
signal-to-noise ratio additional measures have been taken: CCD
images are taken simultaneously with the rf scans during the cooling
cycles and, preceding each measurement, a fast rf scan through the
resonance of the single ion is executed. This ensures single ion
operation and excludes drifts of the resonance due to slow field
fluctuations. During an optical double resonance measurement the
frequency is set alternatingly to seven discrete detunings $\Delta$:
$\Delta=0$~kHz to measure the maximal resonance fluorescence rate
for normalization, $\Delta=\pm$\ 15 kHz to ensure an unchanged
center frequency, $\Delta=\pm$ 250~kHz to measure the background,
and a variable detuning $\pm \Delta$ that is different for each of
these measurement cycles. A measurement cycle for a given value of
$\Delta$ is repeated typically 40 times and finally data points with
equal magnitude of $\Delta$ are averaged and the background is
substracted.

The data-sets with and without field gradient and the corresponding
nonlinear regressions can be compared quantitatively in terms of the
goodness of fit $Q$ as defined in \cite{numrecipes}. While the left
data-set (no gradient) is well described by a single
Lorentzian, quantified by a goodness of fit of $Q=0.95$, the line
profile is substantially altered when a gradient is applied (right
hand side), showing a sideband structure at the axial trap frequency
that is determined independently by parametric heating to
46.0(5)~kHz. A fit with a single Lorentzian (dashed line in figure
5) matches the data-set poorly and gives $Q= 9.6 \cdot 10^{-9}$,
five orders of magnitude below recommended values to reject the
model used for the fit \cite{numrecipes}. Modelling the data by a sum of two
Lorentzians (carrier and sideband) yields good agreement quantified
by $Q=0.92$.

For the relative height of the motial sideband we obtain $a_s / a_0
= 0.084(8)$. From this ratio, $\left< n\right>  = 6.9 (6) \times
10^4$ -- corresponding to a temperature $T=306 (30)$~mK -- is
deduced using equation (\ref{eq:sidebands}) with $\eta_\text{eff}$
independently extracted from the addressing measurements presented
above. This temperature is substantially higher than the Doppler
temperature $T_D$ for Yb ions ($T_{D} \approx 470~\mu$K). It is due
to the delay time of 50 ms between cooling the ion and performing
the measurements, and the low cooling efficiency during the
measurement itself (for 100 ms). Other researchers have observed
similar temperatures with $^{172}$Yb$^+$ \cite{Bell1991,
Kielpinski2006}. For this first demonstration of the coupling
between spin and motional states of trapped ions in a magnetic field
gradient \cite{electrons}, a higher temperature is in fact
advantageous, since it increases the relative height of the motional
sideband. However, for future experiments the heating rate and
temperature of the ions needs to be reduced, though ground state
cooling is not required to obtain high gate fidelities when using
ion spin molecules.

To make full use of the advantages magnetic gradient induced
coupling (MAGIC) and the concept of an  ion spin molecule offers,
substantially higher magnetic field gradients have to be applied in
the future. This will allow for better resolution of individual
ionic resonances and, since then the Rabi frequency may be
increased, too, for fast gate operations. When increasing the number
of ions in a linear trap, their spatial separation decreases, again
making necessary a larger field gradient to separate neighboring
resonances. An expression relating the necessary field gradient for
a given small value of crosstalk to the number of ions and the axial
trap frequency is given in \cite{Mintert2001}. The use of
micro-structured ion traps \cite{Seidelin2006} that are currently under development will
give gradients up to 100 T/m, thus making about 40 Yb ions
distinguishable in frequency space at a trap frequency of $2\pi
\times 260$~kHz. Also, the effective Lamb-Dicke parameter scales
linearly with the field gradient and the spin-spin coupling
constants $J_{ij}$ with the square of the gradient. Thus, coupling
constants useful for QIP and quantum simulations will be attained
with such traps.

Long-lived ground state hyperfine states of $^{171}$Yb$^+$ are well
suited as a qubit for further experiments
\cite{Wunderlich2003,Balzer2006,Olmschenk2007}. Higher Rabi
frequencies (i.e., strong coupling between rf / microwave radiation
and qubit) will be achieved in the future by using micro-structured
elements for generating rf / microwave radiation integrated into an
ion trap.

We thank Chr. Schneider for support in operating the experiment.
Financial support by Deutsche Forschungsgemeinschaft and the
European Union (IP QAP) is acknowledged.

\bibliographystyle{apsrev}

\end{document}